# Genie: A Secure, Transparent Sharing and Services Platform for Genetic and Health Data

Shifa Zhang, Anne Kim , Dianbo Liu , Sandeep C. Nuckchady, Lauren Huang, Aditya Masurkar,
Jingwei Zhang, Lawrence Tseng, Pratheek Karnati, Laura Martinez, Thomas Hardjono , Manolis
Kellis , Zhizhuo Zhang

1.Massachusetts Institute of Technology (MIT), Cambridge, Massachusetts 02142–
1308 Email: fanneykim, dianbo,hardjono , manoli,zhizhuog@mit.edu

2. Secure AI Labs, Cambridge, Massachusetts 02142–1308
Email: fshifa, sandeep, lauren, aditya, jingwei,
larryg@secureailabs.io

3. IBM, 4205 S Miami Blvd, Durham, NC, 27703-9141
Email: karnatip@us.ibm.com

4.Intel Software Guard Extensions, Intel 2200 Mission College Blvd, Santa Clara, CA 95054
Email: laura.martinez@intel.com

*Abstract*—Artificial Intelligence (AI) incorporating genetic and medical information have been applied in disease risk prediction, unveiling disease mechanism, and advancing therapeutics. However, AI training relies on highly sensitive and private data which significantly limit their applications and robustness evaluation. Moreover, the data access management after sharing across organization heavily relies on legal restriction, and there is no guarantee in preventing data leaking after sharing.

Here, we present Genie, a secure AI platform which allows AI models to be trained on medical data securely. The platform combines the security of Intel Software Guarded eXtensions (SGX), transparency of blockchain technology, and verifiability of open algorithms and source codes. Genie shares insights of genetic and medical data without exposing anyone's raw data. All data is instantly encrypted upon upload and contributed to the models that the user chooses. The usage of the model and the value generated from the genetic and health data will be tracked via a blockchain, giving the data transparent and immutable ownership.

## I. INTRODUCTION

### A. Background

Genomics-based personalized medicine began more than ten years ago [6]. Genetic big data has shown promise in conducting breast cancer studies, building the cancer genome atlas (TCGA), and improving screening and diagnosis [37]. Many recent studies have prospective results with advanced machine learning and artificial intelligence (AI) technologies on genotypic and phenotypic big data [23], [9], [38], [15], [25], [17]. Using large amounts of federated genetic and medical data to train AI models and using these models to predict diseases, drug responses, and personality traits will allow for great advancements benefiting human health.

At the same time, the amount of data is growing very fast. Th DNA sequencing has become cheaper, better, and faster in recent years [21], [33]. The Electronic Health Record (EHR) systems are more wildly adopted and generating huge amount of data. However, regulations of both the Health Insurance Portability Accountability Act (HIPAA) in U.S. and General Data Protection Regulation (GDPR) [12] in EU require a strict protection on the private information in the data. Most of the medical data are no accessible to many promising health care AI algorithms because of privacy protection regulations. There is a great need for a secure AI platform for AI models to process on sensitive medical data.

### B. Challenges

The central problem we tackle is how to protect private information and preserve data ownership while sharing infor-mation derived from the data in an open, transparent online environment. Data are easily copied when shared. Once the data are copied, ownerships of the data are eroded. Further-more, there is no way to track data accesses and modifications for those copied data.

Work by Hardjono, Shrier, and Pentland [35] on the open algorithms (OPAL) paradigm points to the need for the sharing of data and insights in a privacy-preserving manner. Additionally, personal data is now recognized as a new asset class [40], which introduces the need for individuals to have the ability to consent to their data being used in computations [12]. There is a clear need for a system that can respect a person's rights to their genetic and health data in order for that data to be accessible to others.

A centralized database could have security facilities to provide a secure environment for data users to access the data without compromising privacy, but these databases are isolated systems largely incompatible with each other, vulnerable to attacks from insiders, and challenging to track once data has been copied to external locations. Therefore, a centralized solution is insufficient.

In this paper, we introduce the Genie (an acronym for Genetic data Exploration by blockchaiN Interconnected Encryption) platform which is an open, distributed, transparent, and secure marketplace to provide high quality genetic and phenotypic big data, AI models, and a secure computation platform, accelerating AI advancing in health care.

## II. EXISTING STUDIES

### A. Homomorphic Encryption

Kim and Lauter introduced private genome analysis through homomorphic encryption [20]. Homomorphic encryption al-lows for computation on encrypted data without needing decryption. The result of the computation is encrypted and can only be decrypted with the same key used to encrypt the input data. It is possible to do statistics or AI model training on homomorphically encrypted user data without decryption. Only computational results are decrypted. This approach protects private information contained in the raw genetic and phenotypic data.

Homomorphic computation was first raised in 1978 [29], but there was little progress until Gentry published his thesis about a fully homomorphic encryption scheme [14]. In 2012, Fan and Vercauteren published an improved homomorphic encryption scheme (FV) based on Gentry's scheme [13], bringing it a step closer to real applications. Some open source homomorphic encryption libraries have been developed based on the FV scheme, such as Microsoft SEAL [8].

Even though the performance of homomorphic encryption and computation has improved significantly in recent years, it is still too expensive to do useful computation for the purposes of genotypic and phenotypic big data analysis, based on performance data from Bajard, et al. [2].

There are some other issues with the homomorphic encryption for private data sharing. First, If allowing arbitrary computations (normally required by AI training) on the encrypted data, it can infer raw data from the computation results. Second, it is difficult to use data from multiple data owners because homomorphic computation requires that all input data are encrypted with the same encryption key.

### B. Hybrid Homomorphic Encryption and Intel SGX [19]

The Secure gwAs in Federated Environment Through a hYbrid solution with Intel SGX and Homomorphic Encryption (SAFETY) framework with hybrid homomorphic encryption and Intel Software Guard Extensions (SGX) was proposed for genome-wide association study (GWAS) research in 2017 by Sadat, et al. [32]. The framework uses homomorphic encryption to encrypt a data owner's data and do basic statistics with homomorphic computation on the encrypted data. Afterwards, the statistics results are sent into SGX private regions of memory, called enclaves, to be decrypted for further computation. Researchers using the enclave can query the results in the enclave. This hybrid framework can get higher performance than pure homomorphic computation because time-consuming multiplying operations can be done in the enclave with the unencrypted data. Because computations on the raw data occur only inside the enclave, this approach still protects the privacy of data owners.

However, if the data user can program the software in the enclave, it is still possible to reverse the computation on the homomorphically encrypted data and extract the raw data. If the user cannot program the software in the enclave, it limits the extent of analysis that can be done on the data.

In the SAFETY framework, data usage is not tracked. Therefore, data owners cannot be compensated for data usage.

### C. Blockchain for Health Care

Kuo, Kim, and Machado introduced blockchain distributed ledger technologies for biomedical and healthcare applications [22]. The blockchain has the advantages of being distributed, robust, tamper-resistant and transparent compared with traditional relational databases. Using the blockchain in biomedical and healthcare fields bring the benefits of improved medical record management, enhanced insurance claim processes, ac-celerated clinical/biomedical research, and advanced biomed-ical/healthcare ledgers. Disadvantages of the blockchain in these fields include too much transparency when handling confidential information, restrictions on speed and scalability, and possible >50% malicious attacks [22].

Linn and Koo introduced blockchain for health data [24]. The authors pointed out that blockchain technology ad-dresses interoperability challenges in health data management. Blockchain is based on open standards and is widely accepted.

## III. PRIMARY PRINCIPLES

This section introduces the primary principles of the Genie platform: privacy-protected and data ownership-preserved data sharing with open algorithm/open source code, Intel Software Guarded eXtensions (SGX), and the blockchain technologies.

### A. Open Algorithm and Source Code

The two techniques that open algorithms (OPAL) [35] use to protect privacy in shared data are as follows:

1) Algorithms are sent to secure and trusted data storages to be evaluated directly on the data instead of sending the private data somewhere else to be processed. The algorithms are publicly inspected and share only results that will never compromise the raw data.
2) The algorithms and each evaluation of the algorithms are logged in an immutable database, such as a blockchain.

In the Genie platform, open source code is an important as-pect of the OPAL paradigm. The source code of the distributed application (Dapp), SGX enclave, and AI model evaluation algorithms are all openly shared online. The cryptographic hashes of these open source codes are registered on the blockchain and can be used by anyone to verify that they have a copy of the original source code.

Open source software has become very popular, 20 years since the book Open sources: Voices from the open source revolution was published [11]. The movement has been driven by not only lower development costs but also better security. Open source software has more eyes looking at it, making it less likely to has security flaws. Users can be confident that open source software will not maliciously access or distribute personal information.

Genie provides a trusted ecosystem of data handling soft-ware. All of the platform's software, including the SGX en-clave software, user-side software, and distributed application (Dapp), that processes raw private data and generates outputs to any human users must be open source. Professional security auditors and the platform users can inspect everything the system does and ensure that it does nothing harmful to their data. The auditors can also re-compile the code to generate the initial state of the software, which can be used to verify the software's installation package.

Open source software also provides several other advantages. The accessibility of open source code allows a large number of people to contribute to ongoing improvements of the software, including the identification of vulnerabilities. One disadvantage, however, is that vulnerabilities in the soft-ware are also open to the public, which could provide an attacker the knowledge to mount attacks on deployments of the software that have not yet been patched.

To minimize the risk of vulnerabilities, we carry out security inspection, testing, and third party auditing for each new release. The platform's users can also inspect the software by themselves because they have the entirety of the source code.

### B. The SGX Trusted Execution Environment

Intel CPU Software Guard eXtensions (SGX) can create secure enclaves. An enclave is a hardware-isolated section of CPU memory which cannot be accessed from outside of the enclave, even with system privileges [10]. An SGX enclave can be used to run secure software and store sensitive data such as passwords, private keys, and personal data.

While an SGX enclave is isolated from the outside world, it is not safe if the software running inside it is malicious. The process of attestation ensures that code running inside an enclave is tamper-free [26], [18], [5]. Basically, the attestation process asks the platform on which the enclave is running to provide proof of the software's initial state in the enclave. Then, the proof is signed by the secret private key of the CPU on the platform. The Intel Attestation Service (IAS) can verify the signature and approve that the software running in the enclave has specific initial state which is the same as the executable image of the enclave.

But there are still two issues even when the enclave has been attested. First, how to know if the software running in the enclave is safe? This question can be answered by the audit reports on the open source codes. And the attestation can ensure the running enclave is identical to the software being audited.

Second, how can the attester know that the public key belongs to secure hardware and not a malicious device? The answer is that the public key must be provided or certified by a trusted organization. For SGX, a secret private key and a public key pair is generated during the manufacturing of the CPU. The secret key is stored and kept secret inside the CPU. The public key is stored by Intel. Intel doesn't publish the CPU public keys. Instead, Intel provides an attestation service to verify the signature of SGX CPU.

Therefore, the trust of the enclave relies on both open source code and Intel. Some people criticize SGX because it requires trust in Intel [34]. However, trusted systems must always rely on a root of trust. Users have to trust the manufacturing of the secure CPU and trust that there are no mistakes when handling the keys. Users have to trust the Public-key Infrastructure (PKI) [1] providing the attestation report certificates. Even if Intel directly published CPU public keys, users would still need to trust the certificates for the public keys signed by Intel.

### C. Blockchain

Blockchain is a distributed public ledger based on crypto-graphic technologies first introduced by Nakamoto in 2008 [30].

Blockchain is a peer-to-peer network without any centralized administration. A new blockchain user or node can be created at any time. Each user account includes a key pair: one private key and one public key. The public key is used as the account ID and the private key is kept secret to prove account ownership and to generate signatures.

Blockchain generates a new block to store the new transactions in a fixed period of time. A proof-of-work mechanism is used to select an account (miner) to create the new block. The miner adds the new transactions, a hash of the previous block, and its signature into the new block and adds it to the blockchain. Data written into the blockchain is incorruptible, because any modification on a block needs the block's miner's private key to generate a new signature. Any changes would also need all the following blocks' miners' private keys in order to update the hashes of the modified blocks in their adjacent blocks. Therefore, even if a particular miner's private key is compromised sometime after the block was mined, it is still not possible to modify the block mined by the miner.

The transactions on public blockchains are transparent to anyone in the world. We can take advantage of this transparency to track usage of a data donor's data, even if it is incorporated into an AI model trained with many data donors' data. Both the acts of donating data to a model and querying a model are recorded as transactions.

A smart contract on a blockchain is a piece of executable code which can be used to define business logic and automate transactions. We developed two smart contracts for the platform to provide data registration, payment escrow, and tokens.

The purposes of using blockchain for the platform are:
1) Storing immutable data. This includes cryptographic hashes of raw data and ownership information. The hashes of the data can be used later to audit the data, namely to prove that the off-chain data has not been modified in an unauthorized manner. The validated genetic and health data, enclave images, enclave source codes, AI model data, and attestation reports are too large to be put on the blockchain and must be stored elsewhere. As such, only the hashes of these data are put on the blockchain and are used later to prove the data are tamper-free. Ownership information includes the account IDs of the owners of the data, models, and the enclave instances. Ownership information is used to send revenue generated from the data services to the

- correct accounts according to the business logic defined in the smart contract.
2) Decentralizing authorizations. This allows users to create anonymous accounts by themselves, helpful for protect-ing the privacy of users.
3) Transparency of transactions. It is important for AI model data owned by many data donors who contribute their data to the model training and the model trainer. Provenance is important for data donors who want to know exactly how their data is being used and for model trainers who want to verify the validity of the data.
4) Business automation and monetary/financial incentives. Blockchain smart contracts support payment escrow and auto redistribution of revenues. This not only reduces transaction costs and delays but also enforces the incen-tive structure for data donors and all other contributors.

Even though blockchain has many advantages such as decen-tralized authorization, transparency, and business automation, it also has limitations we need to overcome.

Firstly, the storage space on the blockchain is very limited and costly. It is not feasible to put complete genetic and health data, AI models, or even registration information on the blockchain.

The data on the blockchain is publicly open. No private information should be put onto the blockchain.

On Genie platform, we use off-chain storages for either large or private data. The private data are stored in the data owner's storage behind a firewall or in SGX secure envi-ronments. Public data are stored on multiple public storages such as GeneTank data storage services, IPFS (InterPlanetary File System), Github, and/or Dropbox at the same time. The integrity of these off-chain data is ensured by the data's cryptographic hashes registered on the blockchain.

Secondly, blockchain mining performance is low. Each new block can only be generated in a fixed period (a few minutes) and each block can only store a limited number of transactions, due to block size and/or computational power restrictions. When there are thousands of transactions happening in a short period of time, the mining delay can be long. Many transactions have to increase their transaction fees to let them be mined earlier. The only way to improve the number of transactions per second (TPS), reduce latency, and reduce transaction fees is to do as much as possible off-chain [27]. We carefully designed the system to minimize necessary blockchain transactions to mitigate the performance and high transaction fee issues of the blockchain.

Thirdly, it is not possible to verify the authenticity of user-contributed data with blockchain alone, and encrypted genotype/phenotype information cannot be verified without decrypting it. The Genie platform uses real world, trustable public key infrastructure (PKI) [3] and trusted SGX enclaves to ensure the data registered on the blockchain are trustworthy. For example, the originality of the genetic and health data registered by the data owner can be verified with a digital signature provided by a data validation SGX enclave. The data validation enclave's integrity can in turn be verified with certificates included in its attestation report, signed by Intel IAS under PKI.

D. Data ownership preservation and privacy protection

Inspired by the ChainAnchor architecture for anonymously registering ownership of the IoT devices on the blockchain [16], we register the attested enclave to the blockchain to ensure unchangeable ownership.

Figure 1 depicts the ownership-preserving and privacy-protecting framework. We open source the source codes of the software that handles sensitive data, so security experts and/or data owners can audit the software to ensure its safety. Intel SGX gives us a secure execution environment (or enclave) in which we can run the audited code. The enclave's CPU hardware prevents any access from outside of the enclave even with system privileges, and the software that runs in the enclave can be attested by the CPU-signed measurement of the software loaded in the memory.

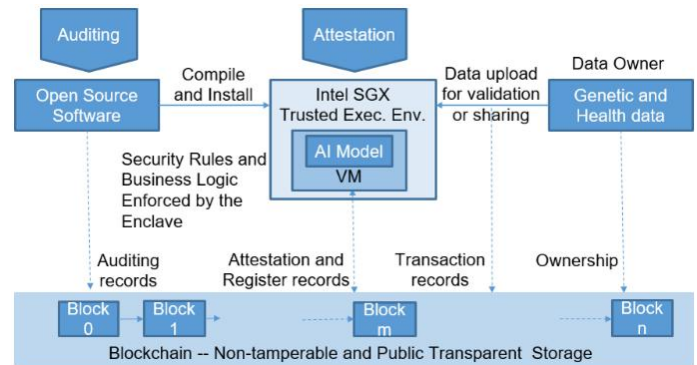

Fig. 1. The Security Framework

The enclave software itself is another important factor for the safety of the enclave. We open source the source codes of the enclave; they must pass security audits before users can trust them.

With source code auditing and enclave attestation, the enclave software becomes a trustable entity which can be used to enforce data safety rules and some business logic when working with the blockchain. These business logic and security rules include:
1) Never send out any data during model training.
2) Delete user data after use.
3) Only accept data validated by a validation enclave.
4) Record every use and payment transaction of AI model prediction services on the blockchain.

Based on the first rule, no human can see the raw data in the enclave. The shared data can only be used by the enclave software to train an AI or statistical model. The query results are generated from the model, not from individual user data. In this way, the insights of genetic and health data are shared without exposing anyone's raw data.

## IV. THE ARCHITECTURE OF THE GENIE PLATFORM

The Genie platform currently is implemented with Ethereum as the blockchain backbone to integrate all the data and services into an anonymous, secure, privacy protected, and open system.

The system is trustless. All the trainers, runners, and users are anonymous participants, none of whom are assumed to be trustful for security. No real-world identification information is required for any of the participants. The openness of the platform allows people to donate data, utilize data, provide services, and use services with very low management costs.

On this trustless platform, the safety of the AI Model and user genetic and prototypic data is well-protected with a combination of open source, blockchain, and secure execution environment technologies as described in the Principles section. The security mechanism of the platform protects all user data (including the AI model algorithms, model parameters, and user genetic and phenotypic data) while these data are being processed, transferred, and stored.

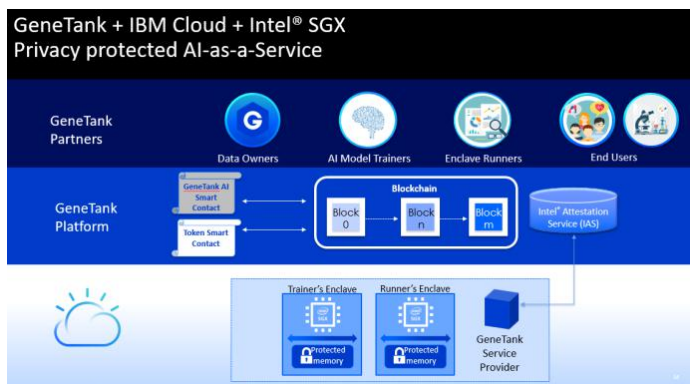

Fig. 2. The Architecture of the Genie Platform

As shown in Figure 2, the platform uses IBM Cloud to host the Genie SGX enclaves powered by Intel processors. Genie users can use the enclaves to train AI models and run the models in a trustable way.

The picture shows the entities of the system as well:

Data owner: Individuals or organizations who own the genetic and health data.

Model trainer: Pharmaceutical companies, biomedical companies, or medical researchers who want to use the data from the data owners to train their artificial intelligence or statistical models.

Enclave runner: People or organizations who have computation resources which can be used to run the enclaves for data validation or model prediction services.

End user: Individuals, hospitals, or pharmaceutical companies who want to use the model prediction services to predict the risk of diseases, drug responses, and traits of individuals with certain genotypes.

Blockchain smart contract: The smart contract developed by GeneTank to register the data, models, and enclaves and automate the business logic.

Intel IAS: The attestation service which is provided by Intel for the enclaves running on Intel CPU platform.

GeneTank Service Provider: A proxy for the enclaves which need to be attested. It forwards the attestation request to Intel IAS, and sends back the attestation reports from Intel IAS.

Enclave: There are three types of enclaves: data validation enclave, model training enclave, and model query service enclave.

1) The data validation enclave validates data from data owners and provide a signature for the data if the data are valid.
2) The model training enclave collects data from data owners and performs model training securely. There is no single output which can be sent out of the enclave.
3) The query service enclave uses genetic and/or phe-notypic data from the end user as input, runs the model to generates prediction results, and sends the results to the end user.

## V. DATA FLOW

To describe how the platform works, we introduce four main data flows of the system in following sections:

A) New Enclave registration, auditing and attestation flow
B) Data preprocessing, validation and registration
C) Model registration and data recruiting
D) AI model query flow.

### A. Enclave registration, auditing, and attestation flow

Enclaves can be developed by anyone who wants to contribute to the platform. The basic requirement for the enclaves is that they must be registered on the blockchain and accepted by the participants of the platform. GeneTank currently develops the enclaves for the initial phase of the platform.

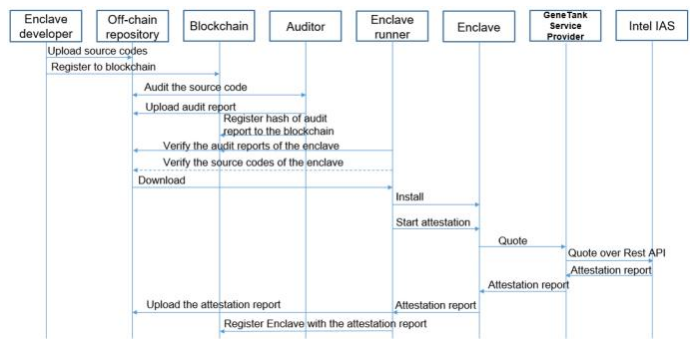

Fig. 3. The Enclave Attestation and Registration Flow

As shown in Figure 3, the enclave developer uploads the source codes, an executable binary image, and description information of the enclave to one or more publicly accessible repositories (e.g. Github, or GeneTank repository), then registers a hash of the source code and a measurement of the executable image of the enclave to the blockchain.

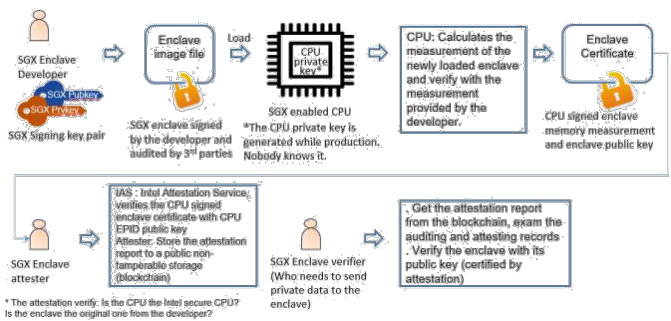

Fig. 4. The Enclave Attestation Flow

The registrations of any information on the blockchain are done with a Distributed Application (Dapp) developed by GeneTank.

The auditors (can be professionals or data owners) audit the source codes of the enclave, and compile the source codes into an executable image, and calculate the measurement of the generated image. The measurement must be the same with the one registered by the enclave developer. The auditor uploads an audit report to the public / off-chain repositories and registers a hash of the audit report on the blockchain.

The enclave runner verifies the enclave by reviewing the audit reports and/or source codes on the repository and checks against the registration records on the blockchain. If the enclave meets all the security requirements, the enclave runner downloads and installs the enclave.

After installation, the enclave runner starts attestation. The enclave generates a quote for itself. The quote includes the measurement of the enclave memory image before initialization, the public key of a self-generated key pair for its identification, some other information, and the signature by the CPU's secret key. The enclave sends the quote to the GeneTank Service Provider server. The server forwards the quote to the Intel IAS through a secure rest API.

The IAS verifies the quote with the EPID of the CPU and generates an attestation report. The report is sent back to the GeneTank Service Provider, then forwarded to the enclave and the enclave runner.

The enclave runner uploads the enclave information including an attestation report, the p2p address of the enclave, and some other information to the repository and registers the hash of the enclave information to the blockchain.

Some details of the enclave attestation flow of the platform are shown in Figure 4.

### B. Data owner registration

When data owners obtain data from DNA sequencing companies or upload phenotypic information by answering questionnaires or providing medical records, they encrypt and upload the data to a data validation enclave for preprocessing and validation.

The validation enclave decrypts data in the secure environment and processes it with an artificial intelligent algorithm to identify fabricated data. Fake data will fail to pass the data validation.

The enclave sends back the processed data and an enclave signed report about the data through an encrypted communi-cation channel. Then the data owner stores the data and the validation report locally and safely and registers the hash of the report to the blockchain.

The registered hash of the report acts as an ownership record of the data and will be used by the model trainer to verify the report. Only the data registered on the smart contract can be used for model training.

The data registration can be withdrawn at any time by the data owner.

This procedure is described in Figure 5.

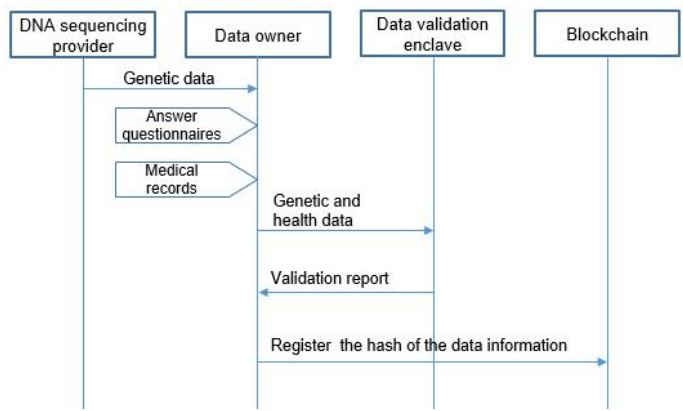

Fig. 5. The Data Validation and Data Owner Registration Flow

Each time the data owners update their data, the data must be sent to a data validation and preprocessing enclave. The data owners will get a data validation report and pre-processed data.

### C. Model registration and data recruiting

A model trainer recruits data from the Genie platform to train a new model. The overall model registration and data recruiting is depicted in Figure 6.

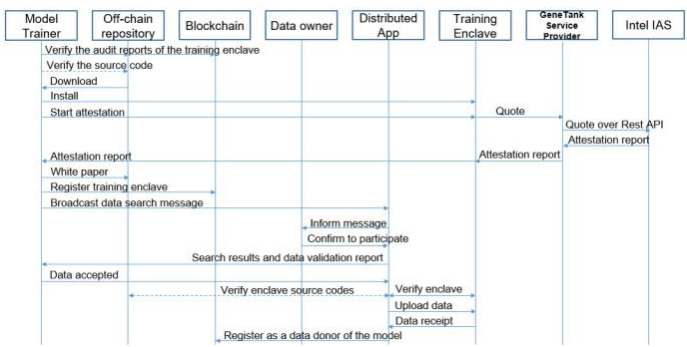

Fig. 6. The Model Trainer Registration and Data Searching

The trainer has to go through the following steps:
Install an enclave package registered on the blockchain

- Get the newly installed enclave instance attested
  - Upload a white paper of the model to the off-chain reposi-tory to describe what service the model will provide, what data are required for the model training, what are the instant payments for the data donors, how the revenues of the model will be shared, etc.
  - Register the model as recruiting model to the blockchain.

After a training model is registered, the data recruiting follows these steps:

- The trainer sends data recruiting messages (p2p message through the blockchain) to all the data owners
- Each Dapp (the same application as for registration) processes the message and analyzes whether the local user data meet the data requirements. If yes, the Dapp informs the data owner to check the conditions and rewards of the model. If the data owner confirms to share the data, the Dapp sends a search result and the validation report of user data to the model trainer. The data owners can also configure their interests' types of models (whitelist) or the types of the models they don't like (blacklist).
- The model trainer checks the search result and the data report. If it is acceptable, the trainer asks the data owner to send the data to the training enclave
- The data owner reviews the auditing reports of the enclave; (optional) redoes the auditing by themselves to further ensure safety of the enclave; verifies the enclave with the enclave public key in attestation report registered on the blockchain and some other information in the off-chain repository; (optional) asks the enclave to redo the attestation if the attestation report is not up to date; if the enclave is safe, sends the data to the enclave provided by the trainer and gets a receipt from the enclave through an encrypted secure channel
- The data owner registers himself as a data donor of the model with the receipt (including quality level, a signature of the enclave)
- The smart contract verifies the receipt from the enclave before it accepts the donor registration.

The data owners may withdraw their data anytime during model training but if the data have been used, the effects of the data in the model may not be removed.

### D. Model training and model runner enclave registration

Model training is done within the enclave. Nobody can see the intermediate and final results of the training. The trainer may use their own data to query the model to evaluate the training result. These evaluation activities must be recorded on the blockchain (enforced by the SGX enclave) as other ordinary queries which must be paid on the blockchain.

Data from the owners are deleted soon after the model is trained to avoid long-term vulnerabilities regarding data safety. This is enforced by the audited codes of the enclave. The donor's data is not possible to be used for training other models unless the data owner decides to participate in these models.

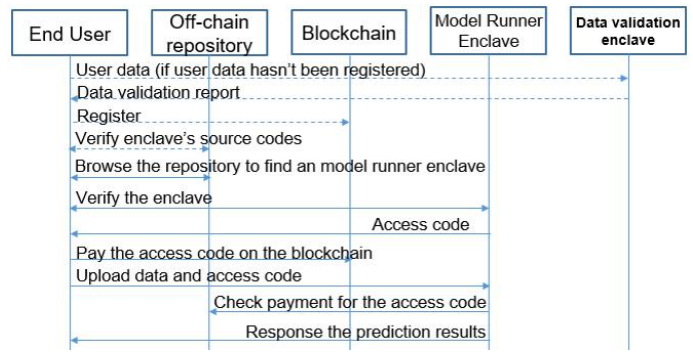

Fig. 7. The Model Query Flow

The trainer changes the status of the model from training model to trained model on the blockchain so that it becomes ready for model running. The trained model is stored inside the enclave. The model trainer registration keeps all the ownership information on the blockchain.

The model running includes the steps of installing model runner enclave, attesting the enclave, and registering the en-clave.

The data flow of the runner enclave installation, attestation, and registration is largely the same as the validation and training enclaves. The differences are that the model codes must be open source and have passed auditing to prevent leaking model data when generating query result outputs.

### E. Model querying

When anyone (end user) wants to use an AI model service to get predictions based on their genetic and health data, they can follow the data flow as shown in Figure 7 through the Dapp. If an end user's data has not been registered yet, they need to validate and register the data to link ownership of that data. The genetic data and an optional part of health data are encrypted and uploaded to the enclave.

The enclave accesses the blockchain to verify a successful payment of the access code. Once the payment is valid, the enclave runs the AI model and returns prediction results. Otherwise, it refuses the request.

After the service is provided successfully, the payment from the end user is transferred to all of the accounts of the model's stakeholders (which include data donors, model trainers, and model runners) or temporarily saved in the smart contract of the blockchain until the stakeholders claim it.

## VI. DETAILS OF DESIGN

We explained how the Genie platform works in previous sections. Some details of the implementation of the plat-form are described below about the P2P communication, blockchain, smart contract, data management, and enclave designs.

### A. P2P communication

The platform is built on a distributed peer-to-peer (P2P) blockchain network which has the advantages of being anony-mous and privacy-protected. The native blockchain can only send broadcasting messages. We extended the P2P protocol to support point to point messages as well. The Dapp can create P2P accounts to receive/send instant messages from/to other platform users. The P2P account is different from the blockchain account which can be found publicly on the blockchain. This can further protect user privacy and avoid unwanted harassment.

The enclaves are also on the P2P communication network. The data owners/donors and the model users can communicate with the enclaves over the P2P network.

Over the P2P network, platform users can:

Search for peers: Search the endpoint (IP addresses and port numbers) of any given P2P account ID;

Send unicast messages: Send messages to a specific P2P account;

Send broadcast messages: Send messages to all the Dapps. The broadcast message sender must be registered on the blockchain as a model trainer. The Dapp can filter the messages according to the preference setting of its user;

Chat in a chat room: A chat room is a channel for a specific topic. The Dapp can join these channels based on the user's interests.

B. Blockchain and Smart Contract

The underlying design eliminates the need for a trustable third party through the use of Ethereum smart contracts [7] and SGX. Smart contracts are symbiotically linked to the blockchain. Writing to a blockchain requires someone to send a transaction which needs to be confirmed by other nodes. The mechanism to achieve this uses either proof of work or proof of stake or a hybrid which is the case in the latest version of Ethereum. Ethereum provides a Turing complete Ethereum virtual machine where smart contracts run. Smart contracts are codes that are executed depending on conditions set within the code. Ether is used to incentivize people to participate and can also be used as a cryptocurrency to facilitate payments in different applications. Each computation in Ethereum has a gas fee.

The Genie platform uses smart contracts to store access control policies [39], [28] like who are registered and whether anybody has tampered with the enclaves, donor's data, and model training parameters. The GeneTank smart contract logs information about data donors, registration information by the creators, trainers and runners, model details trained by the trainer, and any relevant information about access codes generated by the runner for the users. Access codes are used to identify specific user data sent for prediction of a disease. Payments made towards a prediction result of a user's data are distributed to the different parties involved in this process: trainers, runners, and donors. Using tokens as a form of payments is easier than relying on a central authority to distribute fiat currencies.

C. Data management

Private personal data is usually stored locally on a user's device where can protect privacy and ownership very well. For users on mobile devices, they can use the secure data storage service with an SGX enclave on the cloud.

There is also publicly available data such as the source codes and binary codes of enclaves, the white papers for data recruiting, the attestation reports of the enclaves, and the information of the resources available on the platform. These public data have at least one copy which is stored on the GeneTank server. The platform supports other public storages such as Github, IPFS, etc. to improve accessibility.

D. The enclave software and AI model container

The enclave software design is one of the cornerstones of the platform. Security is the first priority but the performance and scalability for biomedical big data processing is also critical for the success of the platform.

We create a virtual machine (VM) as a container for the AI model software which run the AI algorithms. The container isolates the AI model from other parts of the enclave. The AI model running in the container can only access data provided through an input channel. The output from the container is tightly controlled by the enclave software. This design allows the container to run secret AI model training algorithms without fears of compromising the security of the data. It can protect trainer's proprietary AI algorithms and the privacy of data owners at the same time.

The virtual machine currently supports AI algorithms written in the R programming language. R is wildly adopted for both biomedical and deep learning applications. It matches the requirements of both bioinformatics data processing and machine learning very well.

The VM runs the R bytecode compiled with an R compiler [36]. The bytecode programs can be transferred into the enclave and run on the VM so that the enclave can do many different models without changing the code of the enclave. It makes the enclave software more stable and much less auditing work is required.

Any R programs which use the strictly controlled output channel must be open source. R programs with malicious code sending out secret information from the VM can't pass the security audit.

To prevent closed-source AI training model algorithms from sending out private data with side-channel attacks [4], [31], the enclave software randomizes the data input, out-of-enclave memory access, disk operation, inter-enclave communication by inserting dummy operations and rescheduling them to prevent reading out data by monitoring the AI model's external behaviors.

Each enclave has a very limited memory size, much less than the requirements of biomedical big data processing. We developed a virtual memory system using the memory outside of the enclave to expand the memory size in the enclave. All the data written to the external memory or disk are encrypted. The encryption key for external memory is generated each time the enclave starts up. The files permanently stored on the disk are encrypted with an SGX Sealing key [18].

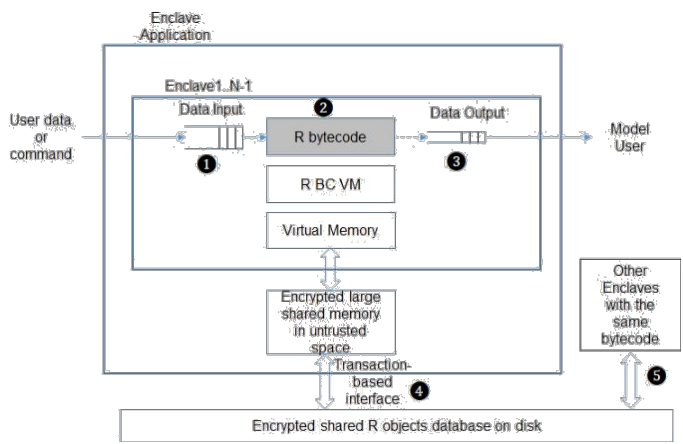

Fig. 8. The Enclave Design with Virtual Machine

We also support scale-out parallel processing which can divide the tasks into smaller ones and allocate to multiple enclaves to process. The enclaves can work in a coordinated manner to improve the performance of training. The enclaves can be deployed on one or multiple computers.

Private data in the enclave can be stored on permanent storage such as hard drives securely with the SGX data sealing feature. The sealed data are encrypted and can only be decrypted by the enclave which saved them. The design of the enclave with virtual machines is shown in Figure 8.

Some explanation about the numbered items in the figure:

1) The inputs include the data from data owners or the query users or commands from the trainers. The depth of the input buffer is randomly changed to prevent side-channel attack from malicious AI Model training R bytecode.
2) The R bytecode is driven by the data inputs include the commands. The bytecode is a blackbox (close source) for training and a whitebox (open source) for querying.
3) Only the querying bytecode which has been registered on the blockchain can generate outputs. The output is encrypted and can only be decrypted by the model query end user.
4) The external disk is a database of R objects. Any read-modify-write operations are managed within transactions to protect data in the event that the system crash.
5) The database can be shared with other enclaves with the same R bytecodes to support parallel training

## VII. CONCLUSION

Genie is a AI-as-a-service marketplace platform, empower-ing by public genetic and health data. Genie provide high-est standard of secure and immutability with Intel SGX, blockchain, and open source technologies to protect the pri-vacy and preserve ownership of data.

The platform is decentralized and open to all individuals or organization data owners. Pharmaceutical companies, biotech companies, and biomedical researchers can use the platform to recruit data easily through distributed data searches for AI model training and create powerful models for disease predictions, drug responses, and personal traits. Individuals and hospitals can access the services provided by these AI models.

The platform enforces the rights of data ownership including possession, control, distribution, and disposal of the owner's genetic and health data. The platform also maintains the data donors' and model trainer's ownership of the AI models. The revenues generated by the AI models' prediction services are transferred automatically to all the owners of the models and the people who run the models. It encourages all the participants to continue contributing to the platform and forms a positive incentive loop.

## VIII. ACKNOWLEDGEMENTS

Bixin Zhang who offered the ideas of P2P communications and edited the text.

This paper was supported by IBM cloud with Intel SGX cloud servers. MIT Sandbox program provides start-up fund for GeneTank. Creative Destruction Lab provides mentorship support.